\documentclass[preprint2,flushrt]{aastex}
\usepackage{txfonts}
\usepackage{subfigure}
\usepackage{color}                       
\usepackage{url}                         
\usepackage{lineno}
\usepackage{amsbsy}
\usepackage{xspace}
\usepackage{wtmmPkg}

\slugcomment{Version of April 2nd, 2010.}
\shorttitle{Characterising Complexity in Solar Magnetogram data}
\shortauthors{Kestener et al.}

\begin{document}

\title{Characterising Complexity in Solar
    Magnetogram Data using a Wavelet-based Segmentation Method}

\author{P.~Kestener\altaffilmark{1},
  P.A.~Conlon\altaffilmark{2},
  A.~Khalil\altaffilmark{3},
  L.~Fennell\altaffilmark{2},
  R.T.J.~McAteer\altaffilmark{2},
  P.T.~Gallagher\altaffilmark{2} and
  A.~Arneodo\altaffilmark{4,5}}
\affil{$^{1}$ CEA, Centre de Saclay, DSM/IRFU/SEDI, 91191 Gif-sur-Yvette, France\\
  $^{2}$ School of Physics, Trinity College Dublin, Dublin 2, Ireland\\
  $^{3}$ Department of Mathematics \& Statistics, The University of
  Maine, Orono, ME, USA 04469\\
  $^{4}$ Universit\'e de Lyon, F-69000, Lyon, France\\
  $^{5}$ Laboratoire Joliot Curie and Laboratoire de Physique, Ecole
  Normale Sup\'erieure de Lyon, 46~all\'ee d'Italie, 69364 Lyon c\'edex 07,
  France
}
\email{pierre.kestener@cea.fr}

\begin{abstract}
The multifractal nature of solar photospheric magnetic structures are
studied using the 2D wavelet transform modulus maxima (\WTMM{})
method. 
This relies on computing partition functions from the wavelet
transform skeleton defined by the \WTMM{} method. 
This skeleton provides an adaptive space-scale partition of the
fractal distribution under study, from which one can extract the
multifractal singularity spectrum. 
We describe the implementation of a multiscale image processing segmentation procedure based on the partitioning of the \WT{} skeleton which allows 
the disentangling of the information concerning the multifractal properties of active 
regions from the surrounding quiet-Sun field. The quiet Sun exhibits a
average H\"older exponent $\sim -0.75$, with observed multifractal properties
due to the supergranular structure. On the other
hand, active region multifractal spectra exhibit an average H\"older
exponent $\sim 0.38$ similar to those found when studying experimental
data from turbulent flows. 
%
\end{abstract}


%
\section{Introduction}
\label{Introduction} 

Since the late 70's and the propagation of fractal ideas throughout 
the scientific community~\citep{bMan82}, 
there have been numerous applications of the concepts of scale 
invariance, self-similarity, long-range dependence in many areas of physics,
chemistry, biology, geology, meteorology, economy, social and material 
sciences~\citep{b89b,bWes90,b94,bFractal94,b95,b95bis,bFri95,bArn95,bDis02}.
 Various methods were developed to quantify scale-invariance properties 
through the computation of the fractal dimension $D_F$ for self-similar 
objects or the roughness
 exponent $H$ for self-affine 
fractals~\citep{bMan82,b87,bFed88,aArg90,aLea93,b95,aTaq95}.
Unfortunately $D_F$ and $H$ are global quantities that do not account for 
the possibility of point-to-point fluctuations of the scaling properties 
of a fractal object.
The multifractal formalism was introduced in the 
mid-eighties to provide a
statistical description of the fluctuations of regularity of singular measures
 that are
found in chaotic dynamical systems~\citep{aHas86,aCol87,aRan89} or in modelling
of the energy cascading process in turbulent 
flows~\citep{aMan74b,aPal87,bMan89,aMen91}.
Box-counting and correlation algorithms were successfully adapted to resolve
 multifractal scaling for isotropic self-similar fractals by computation of the
generalized fractal dimensions $D_q$~\citep{aGras83a,aGras83b,aGra88}.
As to self-affine fractals, Parisi and Frisch~\citep{aPar85} proposed,
for the analysis of fully-developed turbulence velocity data, an
alternative multifractal description based on the investigation of the
scaling behavior of the so-called structure
functions~\citep{bFri95,bMon75}: $S_p(l)=<(\delta v_l)^p>\sim
l^{\zeta_p}$ ($p$ integer $> 0$), where $\delta v_l(x)=v(x+l)-v(x)
 \sim l^{h(x)}$ is an increment of the recorded signal over a distance $l$.
 Then, after reinterpreting the roughness exponent as a local
quantity~\citep{aPar85,aMuz91,aMuz94,aArn95}: $\delta
v_l(x) \sim l^{h(x)}$ (power-law behavior), the $D(h)$ 
{\em singularity spectrum} is defined as
the Hausdorff dimension of the set of points $x$ where the local
roughness (or H\"older) exponent $h(x)$ of $v$ is $h$.  In principle,
$D(h)$ can be attained by Legendre transforming the structure function
scaling exponents $\zeta_p$~\citep{aPar85,aMuz91,aMuz94,aArn95}.
Unfortunately, as noticed by \citep{aMuz91,aMuz93,aMuz94}, both the
  box-counting and structure function methodology have intrinsic
  limitations and fail to fully characterize the corresponding
  singularity spectrum since only the strongest singularities are a
  priori amenable to these techniques. As such, both methods are
  limited in their application to real data sets \citep{aMuz94,Georgoulis:2005,Conlon:2008}.

In previous work, Arneodo and collaborators~\citep{aMuz91,aMuz93,aMuz94,aArn95}
 have shown that there exists a natural way of performing a unified multifractal
 analysis of both singular
measures and multi-affine functions, which consists in using the
 \textit{continuous wavelet 
transform}~\citep{aGou84,aGros84,bMey90,bDau92,bMal98}.
%
By using wavelets instead of boxes, one can take advantage of the freedom 
of the choice of these ``generalized oscillating boxes'' to get rid of possible
smooth behavior that might either mask singularities or perturb 
the estimation of their strength $h$. 
The other fundamental advantage of using wavelets is that the skeleton defined
by the \textit{wavelet transform modulus maxima} (\WTMM{})~\citep{aMalZho92,aMalHwa92},
provides an adaptative space-scale partitioning from which one can extract
the $D(h)$ singularity spectrum \textit{via} the scaling exponent $\tau(q)$
of some partition functions defined from the \WT{} skeleton.
 The so-called \WTMM{}
method~\citep{aMuz91,aMuz93,aMuz94,aArn95} therefore gives access to the entire
 $D(h)$ spectrum \textit{via} the usual Legendre transform 
$D(h)=min_q(qh-\tau(q))$. We refer the reader to \citet{aBac93} and \citet{aJaf97b} for
rigorous mathematical results and to \citet{aHen94} for the theoretical 
treatment of random multifractal functions.

Applications of the \WTMM{} method to \UD{} signals have already
provided insight into a wide variety of problems~\citep{refaarn02},
e.g. the validation of the log-normal cascade phenomenology of fully
developed turbulence~\citep{aArn98,aArn99,aDelArn01} and of
high-resolution temporal rainfall~\citep{aVenug06,aVenug06b,aRoux09},
the characterization and the understanding of long-range correlations
in DNA sequences~\citep{aArn95b,aArn96b,aAud01,aAud02}, the
demonstration of the existence of a causal cascade of information from
large to small scales in financial time series~\citep{aArn98e,aMuz00},
the use of the multifractal formalism to discriminate between healthy
and sick heartbeat dynamics~\citep{aIva96,aIva99}, the discovery of a
Fibonacci structural ordering in \UD{} cuts of diffusion limited
aggregates~\citep{aArn92,aArn92d,aArn92c,aKuh94} and in hard X-ray emission from solar flares~\citep{McAteer2007-apj}. The \WTMM{}
 method has been generalized from \UD{} to \DD{} with the specific
 goal to achieve multifractal analysis of rough
 surfaces~\footnote{The fractal dimension of a rough surface
   associated to the graph $z=S(x,y)$, where $S(x,y)$ 
   represents the height of the surface at location $(x,y)$, is a
   quantity $D_F$ between 2 and 3.} with
 fractal 
 dimension $D_F$ anywhere between 2 and
 3~\citep{aArr97,aArnDec00a,aArnDec00b}. The \DD{} \WTMM{} method has been
 successfully applied to characterize the intermittent nature of
 cloud structure from satellite images~\citep{aArn99c,aArnDec00c} and
 to assist in the diagnosis of breast tissue lesions in digitized
 mammograms~\citep{akes01}. In 
 astrophysics, this method was adapted and used to characterize the
 anisotropic structure of atomic hydrogen gas (HI) in the Galatic
 disk~\citep{kha06}. From the analysis of very large mosaics taken
 from the Canadian Galatic Plane Survey~\citep{tay03}, directional
 roughness exponents were introduced to show that the HI in the
 Galactic spiral arms has a scale-dependent anisotropic signature
 while the HI in the inter-spiral arm regions exhibits
 scale-independent anisotropy. Along that line, the \DD{} \WTMM{}
 method was further applied to characterize the space-scale nature of
 anisotropic structures~\citep{aSnow08a,aSnow08b,akha09} and to perform
 objective segmentation of image features of interest from noisy
 backgrounds~\citep{kha07,acad07,athi09}. We refer the reader to
 \citet{bArn02} for a review of the \DD{} \WTMM{} methodology,
 from the theoretical concepts to experimental applications. Recently,
 the \WTMM{} method has been further extended to \TD{}
 scalar~\citep{aKes03_prl} as well as \TD{}
 vector~\citep{akes04prl,aSerra07} fields analysis and applied to
 \TD{} (velocity, vorticity, dissipation, enstrophy) numerical data
 issued from direct numerical simulations (DNS) of incompressible
 Navier-Stokes equations. Because it combines singular-value
 decomposition and multifractal description, the so-called tensorial
 wavelet transform modulus maxima method for vector fields~\citep{akes04prl,aSerra07}
 looks very promising for future simultaneous multifractal and
 structural (vorticity sheets, vorticity filaments) analysis of
 turbulent flows.

Our aim here is to exploit the ability of the \WTMM{} method to study
compound systems that display some non-analyticity in their
multifractal spectra as the signature of some phase transition between
two underlying scale invariant components with different multifractal
properties~\citep{bBoh88,aMuz94,aArn95}. These two components can both
have some physical significance as previously experienced when using
the \WTMM{} method to detect vorticity filaments in swirling turbulent
flows~\citep{aRouMuz99} or microcalcifications from breast tissue
background in digitized mammograms~\citep{akes01,bArn02}. One of these
components can be noise that may cause drastic distortions in the
returned multifractal spectra. In this work we will follow a
wavelet-based strategy inspired from the one previously used in \UD{} to detect
replication origins and promoters as jumps (discontinuities) in \UD{}
noisy skew profiles in mammalian
genomes~\citep{aBro05,touchon05,nicolay07} and in \DD{} to
perform an objective and automatic segmentation of chromosome
territories in fluorescence microscopy imaging of mouse cell
nuclei~\citep{kha07,acad07} and of gold formation on vapodeposited
thin gold films~\citep{athi09}.

The purpose of the manuscript is to demonstrate the suitability and
reliability of the \WTMM{} method to propose a wavelet-based
segmentation procedure adapted to solar magnetogram data. 
In section~\ref{segmentation},
the basics of the \DD{} \WTMM{} method are presented. Its ability to
disentangle the underlying scale invariant components of a compound
system displaying a phase transition in its singularity spectra is
discussed and a strategy of segmentation is
implemented. Section~\ref{section_synthetic_data} is devoted to a test
application of the proposed segmentation procedure on a theoretical data
set with known multifractal properties. In section~\ref{observations}
we report the results obtained when using this wavelet-based
segmentation method to separate active regions from quiet-Sun features
in solar line-of-sight magnetogram data. Our conclusion and future
directions are then given in Section~\ref{conclusions}.

%
\section{Segmentation methodology of compound multifractal systems
  using the \DD{} \WTMM{} method}\label{segmentation}

\subsection{Basics of the \DD{} \WTMM{} method}
\label{wtmm_method}

The main steps of the \DD{} \WTMM{} method are presented here.
Details can be found in \citet{aArnDec00a,aArnDec00b,bArn02,kha06}.

\begin{enumerate}
\item Computation of the \DD continuous wavelet transform of the input
 image function $f({\bf x})$ with analyzing wavelets defined as the partial
 derivatives of a smoothing Gaussian kernel $\phi$:
\begin{eqnarray}
  {\bf T}_{{\bf \psi}}[f]({\bf b},a)\nonumber
  & =&\Bigg( \matrix{ T_{\psi_1}[f]
    = a^{-2} \int d^2 {\bf x}  \; \psi_1 \big(a^{-1}({\bf x} - {\bf
      b})\big) f({\bf x}) \cr T_{\psi_2}[f] = a^{-2} \int d^2 {\bf x}
    \; \psi_2 \big(a^{-1}({\bf x} - {\bf b})\big) f({\bf x}) } \Bigg),
  \nonumber\\[0.3cm]
  & = &\nabla \{ T_{\phi}[f]({\bf b},a)\} \\
  & = &\nabla \{ \phi_{{\bf b}, a} * f \},\nonumber
 \label{eq_wtmm_1}
\end{eqnarray}
where $\psi_1 = \partial\phi/\partial x$, $\psi_2
= \partial\phi/\partial y$ and $\phi({\bf x})=\exp(-|{\bf
  x}|^2/2)$. 
Eq.~(1) amounts to define the \DD{} wavelet
transform as the gradient vector of $f({\bf x})$ smoothed by a dilated
version $\phi(a^{-1}{\bf x})$ of the Gaussian filter.
\item For each scale $a$, extract the \WTMM{} edges defined as the
 locations ${\bf b}$ where the \WT{} modulus $\Mpsi [f]({\bf b}, a)
   = | \Tpsi [f]({\bf b},a)|$ is locally maximum in the direction of
   the \WT{} vector $\Tpsi [f]({\bf b},a)$. These \WTMM{} points lie
   on connected \textit{maxima chains}. Along each of these maxima
   chains, locate the local maxima called \WTMMM{} for \WTMM{}
   maxima. Note that the two ends of an open maxima chain are not
   allowed to be a possible \WTMMM{} location.
\item Extract the \WT{} skeleton which is the set of  \textit{maxima
 lines}  ${\cal L}_{{\bf x}_0}$ obtained by connecting these \WTMMM{}
from scale to scale, starting at location ${\bf x}_0$ at smallest
scale. 
Start at the smallest scale $a_{min} \sim 7 pixels$ (minimum size of
the support of the wavelet function) and 
link each \WTMMM{} to their nearest neighbor found at the scale just
above. Proceed iteratively from scale to scale up to the largest scale
$a_{max}$ (limited by the image size and border effects in wavelet
transform computations). It is important to recall here that these lines contain all
the information about the local H\"older regularity properties of the
function $f$ under consideration and that along a maxima line ${\cal
  L}_{{\bf x}_0}$ that points to ${\bf x}_0$ in the limit $a \to 0^+$,
the wavelet transform modulus behaves as a power law:
\begin{equation}
 {\cal M}_{\bf \psi}[f][{\cal L}_{{\bf x}_0}(a)] \sim a^{h({\bf x}_0)},
 \label{eq_wtmm_2}
\end{equation}
where $h({\bf x}_0)$ is the H\"older exponent, \textit{i.e.} the strength of
the singularity of the function $f$ at the point ${\bf x}_0$.
\item From the \WT{} skeleton compute the partition functions:
\begin{equation}
  {\cal Z}(q,a) = \sum_{{\cal L} \in {\cal L}(a)}  \left[{\cal M}_{{\bf \psi}} [f] ({\bf x \in {\cal L}},a)\right]^q,
\label{eq_wtmm_3}
\end{equation}
which allows to define the $\tau(q)$ scaling exponents as
\begin{equation}
{\cal Z}(q,a) \sim a^{\tau(q)}, \;\; a \to 0^+.
\label{eq_wtmm_4}
\end{equation}
One can optionnally compute the companion partition functions $h(q,a)$
and $D(q,a)$ and define the corresponding scaling exponents when $a \to 0^+$
\begin{eqnarray}
h(q,a) =
 &\sum_{{\mathcal L}\in{\mathcal L}(a)} \ln \left|
 \ron{M}_{\bpsi}[f]({\mathbf x},a) \right| \;
 W_{\bpsi}[f](q,{\mathcal L}, a)\; \sim a^{h(q)}, \label{eq_wtmm_5}\\
D(q,a) =
 &\sum_{{\mathcal L}\in{\mathcal L}(a)}
 W_{\bpsi}[f](q,{\mathcal L}, a) \; \ln 
 \bigl( W_{\bpsi}[f](q,{\mathcal L}, a) \bigr) \; \sim a^{D(q)}.\label{eq_wtmm_6}
\end{eqnarray}
where 
\begin{equation}
W_{\bpsi}[f](q,{\mathcal L}, a) = \left[ \ron{M}_{\bpsi}[f]({\mathbf
  x},a) \right]^q/{\cal Z}(q,a)
\label{eq_wtmm_7}
\end{equation}
\item Compute the $\tau(q)$ spectrum by performing linear regression
  fits of $\ln {\cal Z}(q,a)$ vs $\ln a$ and finally compute the
  $D(h)$ singularity spectrum by Legendre transforming $\tau(q)$:
  \begin{equation}
    D(h) = \min_q \left[ qh-\tau(q)\right].
    \label{eq_wtmm_8}
  \end{equation}
Alternatively $D(h)$ can be computed from the estimate of the scaling
exponents $h(q)$ and $D(q)$ in Eqs. (\ref{eq_wtmm_5}) and
(\ref{eq_wtmm_6}) respectively.
\end{enumerate}
Note that alternative approaches to the \WTMM{} method have been
developed using discrete wavelet bases including the recent use of
wavelet leaders~\citep{bJaff06,aWen07a,aWen07b}. We think that the
continuous \WT{} better 
suits our goal to provide a selective multifractal analysis of
multi-component images via some objective segmentation of maxima lines
in the \WT{} skeleton.

\subsection{Adapting the \DD{} \WTMM{} method to the
  segmentation of compound multifractal systems}
\label{subsection_wtmm_segmentation}
\begin{figure}
\plottwo{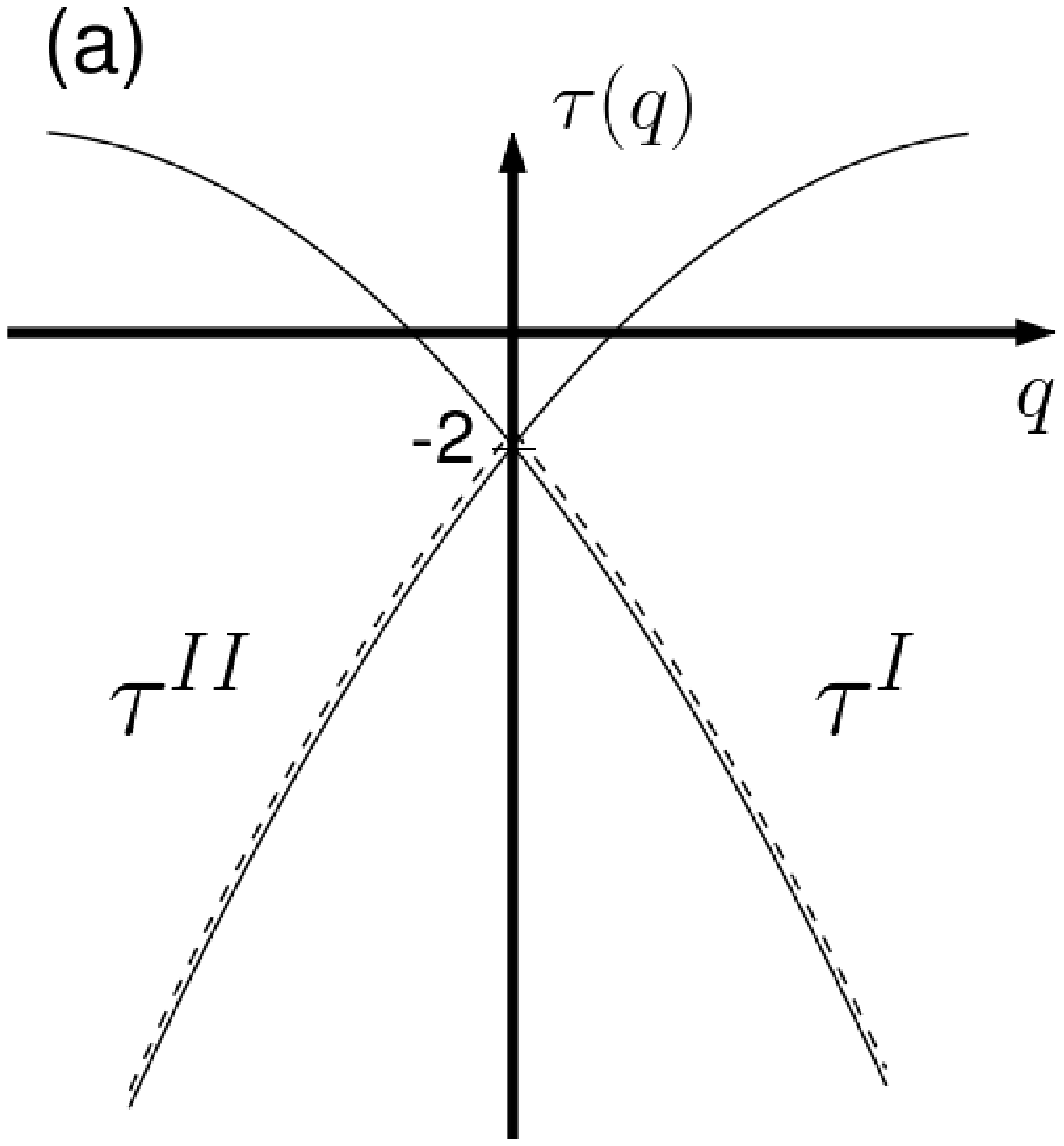}{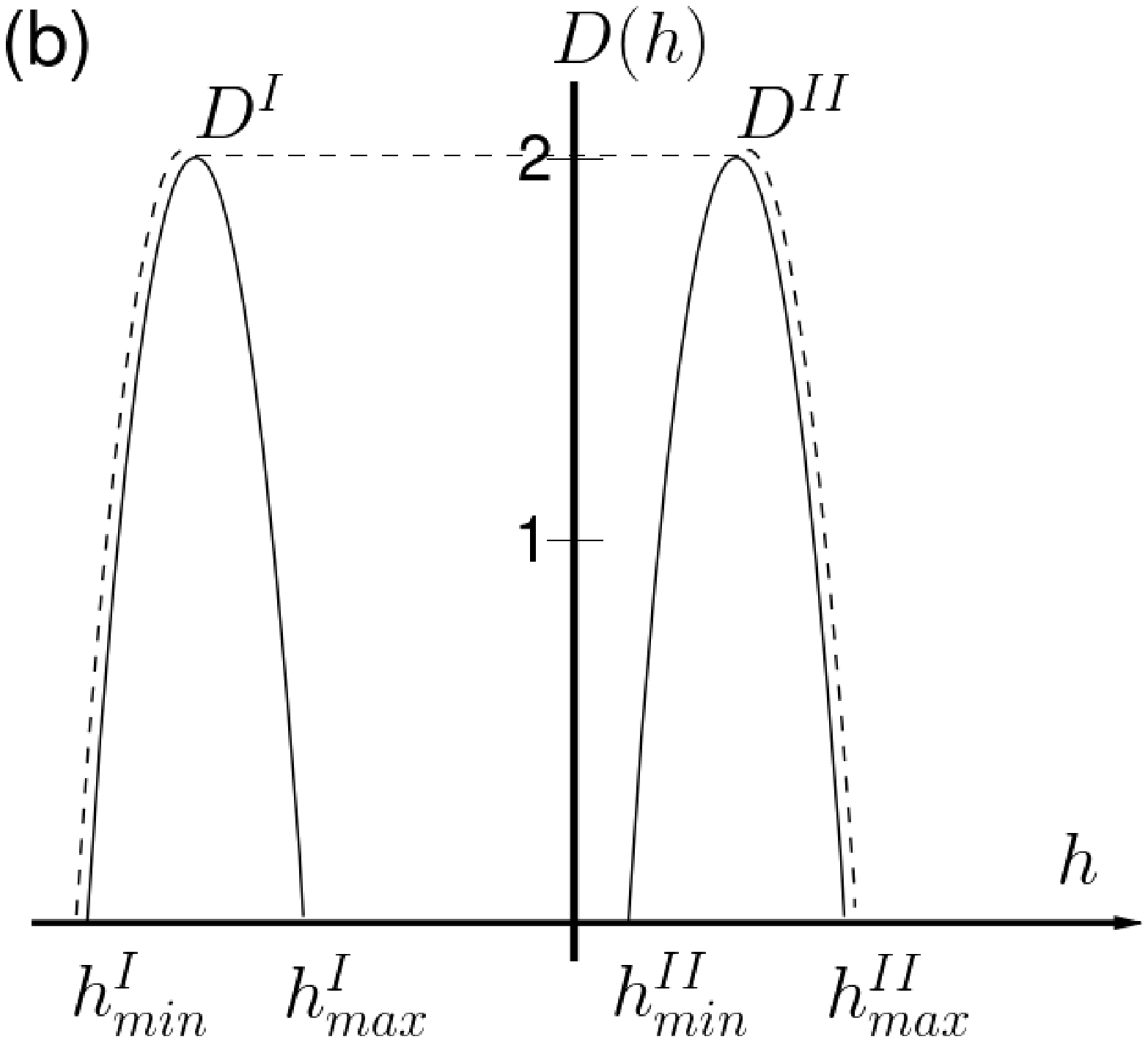}
\caption{Illustration of a phase transition in the multifractal
  spectra of a compound system (Eq.~(\ref{eq_wtmm_9})). $(\tau^{I}(q),
  D^{I}(h))$ and $(\tau^{II}(q),D^{II}(h))$ are the multifractal
  spectra for $f^{I}({\bf x})$ and $f^{II}({\bf x})$ respectively. The
  \DD{} \WTMM{} method and more generally the multifractal formalism,
  give access to the dashed $\tau(q)$ curve (Eq.~(\ref{eq_wtmm_11})) in
  (a) and via the Legendre transform (Eq.~(\ref{eq_wtmm_8})) to the
  dashed $D(h)$ sprectrum in (b) which is the supremum of the
  $D^{I}(h)$ and $D^{II}(h)$ spectra.}
\label{fig_tqdh_theorique}
\end{figure}

For simplicity, we will assume that the compound multifractal systems
of interest here can be considered as the sum of two scale invariant
components:
\begin{equation}
f({\bf x}) = f^{I}({\bf x}) + f^{II}({\bf x})
\label{eq_wtmm_9}
\end{equation}
characterized by the singularity spectra $D^{I}(h)$ and $D^{II}(h)$
respectively. Ideally we will further  suppose that $D^{I}(h)$ and
$D^{II}(h)$ have non-overlapping support $[h^{I}_{min}, h^{I}_{max}]
\cap [h^{II}_{min}, h^{II}_{max}] = \emptyset$ and that $h^{I}_{max} <
h^{II}_{min}$ meaning that $f^{I}({\bf x})$ possesses stronger
singularities than $f^{II}({\bf x})$. In the limit $a \rightarrow
0^{+}$, the partition function ${\cal Z}(q,a)$ (Eq. (\ref{eq_wtmm_3}))
can be split into two parts:
\begin{equation}
  {\cal Z}(q,a) = {\cal Z}^{I}(q,a) + {\cal Z}^{II}(q,a) =
  C_{I}(q)a^{\tau^{I}(q)}+C_{II}(q)a^{\tau^{II}(q)},
\label{eq_wtmm_10}
\end{equation}
where $C_{I}(q)$ and $C_{II}(q)$ are prefactors that depend on
$q$. Since $h^{I}_{max} < h^{II}_{min}$, it follows easily that in
this limit, there exists a critical value $q_{crit}$ so that:
\begin{equation}
  \tau(q) = \left\{
    \begin{array}{lll}
      \tau^{I}(q) & \mathrm{for} &q>q_{crit}\\
      \tau^{II}(q) & \mathrm{for} &q<q_{crit}\\
    \end{array}
  \right.
\label{eq_wtmm_11}
\end{equation}
Therefore, the $\tau(q)$ spectrum has a non-analyticity at $q_{crit}$;
when crossing this critical value, there is a transition
from one scale invariant component to the other. 
As illustrated in Fig.~\ref{fig_tqdh_theorique}, when Legendre
transforming Eq.~(\ref{eq_wtmm_11}), one gets the upper envelop of the
$D^{I}(h)$ and $D^{II}(h)$ spectra, a classical result for entropy in
equilibrium statistical physics~\citep{bBoh88,aMuz94,aArn95}. In that
respect the classical \DD{} \WTMM{} method does not provide separate access
to $D^{I}(h)$ and $D^{II}(h)$.

Our segmentation strategy consists in using the \WT{} skeleton
to discriminate the maxima lines ${\cal L}^{I}(a)$ associated with
singularities of $f^{I}({\bf x})$ and maxima lines ${\cal L}^{II}(a)$ associated with
singularities of $f^{II}({\bf x})$, over the range $[a_{min},a_{max}]$ of accessible
scales. This can be done theoretically by
comparing the power-law behavior of ${\cal M}_{\bf \psi}[f]({\bf x}
\in {\cal L})$ along each maxima line of the \WT{} skeleton
(Eq.~(\ref{eq_wtmm_2})).
From the local estimate of $h({\bf x})$, one can
expect to partition the \WT{} skeleton into two sub-skeletons, one made of
the maxima lines ${\cal L}^{I}(a)$ and the other one made of maxima
lines ${\cal L}^{II}(a)$. In practice, this partitioning will suffer
from the finite range of scales available to the analysis and the
desired segmentation will require special care as far as finite-size
effects and statistical convergence issues are concerned.

%
\section{Application of the wavelet-based segmentation method to
  synthetic data}
\label{section_synthetic_data}

\begin{figure}
\centerline{\includegraphics[width = 0.65\linewidth,angle=-90]{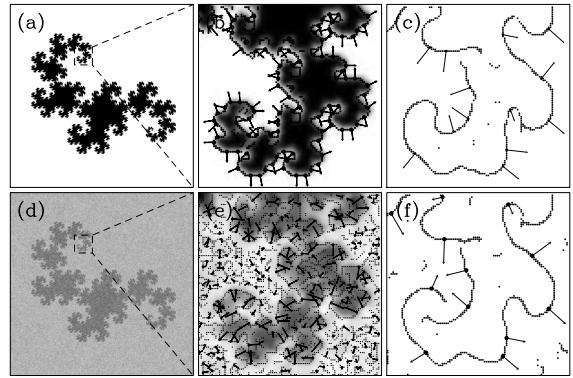}}
\caption{(a) The fractal Dragon ($1024 \times 1024$). (b) \WTMM{}
  chains at the smallest scale $a=\sigma_W = 7$ pixels
 in a small $100\times 100$
 region of the fractal Dragon. (c) Same as in (b) for scale $a=2\sigma_W$. (d) The
 fractal Dragon embedded in a fractionnal Brownian noise
 ($H=-0.7$) background of
 twice as large amplitude. (e) and (f) are the same as (b) and (c) but
 for the noisy fractal Dragon. In (b,c,e,f), the black arrows
 represent the \WT{} vectors originating at the \WTMMM{}.
 In (b) and (e) the background image is the
 smooth-convoluted image $\phi_{{\bf b}, a} * f $ at scale
 $a=\sigma_W$.}
\label{fig_dragon}
\end{figure}
We consider an academic example image with two fractal components:
the fractal Dragon~\citep{duda-2007}
embedded into a noisy background generated from fractional Brownian noise with
H\"older exponent $H=-0.7$.
The fractal Dragon is a self-similar fractal defined as the limit set of
an iterated function system (the Lindenmayer system) of the same type
as the one used to generate the Sierpinski gasket or the Von Koch curve~\citep{bMan82},
but the fractal Dragon has less obvious geometrical symmetries. Let us
note that the fractal Dragon is space-filling, meaning its
fractal dimension is 2, whereas its boundary has a fractal dimension
known analytically 
$D_{Dragon}=\log_2(\frac{1+\sqrt[3]{73-6\sqrt{87}}+\sqrt[3]{73+6\sqrt{87}}}{3})
\simeq 1.5236$. 
A sample fractal 
Dragon is shown in Figure \ref{fig_dragon}(a) whereas the corresponding noisy
two-component image is shown in Figure~\ref{fig_dragon}(d).
As previously discussed, the
wavelet analysis proposed in this work is sensitive to singularities,
\textit{i.e.} to points in the images where the signal is singular. We expect
the \WTMM{} analysis of the image shown in Fig. \ref{fig_dragon}(d)
to simultaneously reveal multifractal information about both the
boundary of the fractal Dragon  
and of the rough background texture.
Let us recall that the two
components have known mono-fractal type self-similar properties,
\textit{i.e.} a singularity spectrum degenerated to a single point: ($h=-0.7$,
$D=2$) for the fractional Brownian noise and ($h=0$,
$D=D_{Dragon} \simeq 1.5236$) for the boundary of the fractal Dragon. The roughness
$H=-0.7$ of the fractional Brownian noise was chosen to mimic the
texture of the quiet-Sun images (see Section \ref{sect_QS}).
Figures \ref{fig_dragon}(b) and \ref{fig_dragon}(e) illustrate the results of the
computation of the 
\WT{} maxima chains at the smallest scale; the arrows correspond to
the \WT{} vectors (Eq.~(\ref{eq_wtmm_1}))
at the \WTMMM{}
locations. Figures \ref{fig_dragon}(c) and \ref{fig_dragon}(f) show the maxima chains
at a scale twice as large as in Figures \ref{fig_dragon}(b) and \ref{fig_dragon}(e).
When going from large to small scales, whereas the boundary of the
fractal Dragon is better and better approximated by some \WTMM{}
chains (edge detection in the smoothed image),
an increasing
number of additional maxima chains start emerging as the signature of
the presence of a colored noise (Fig.~\ref{fig_dragon}(e) as compared
to Fig.~\ref{fig_dragon}(b)).

\begin{figure}
\plottwo{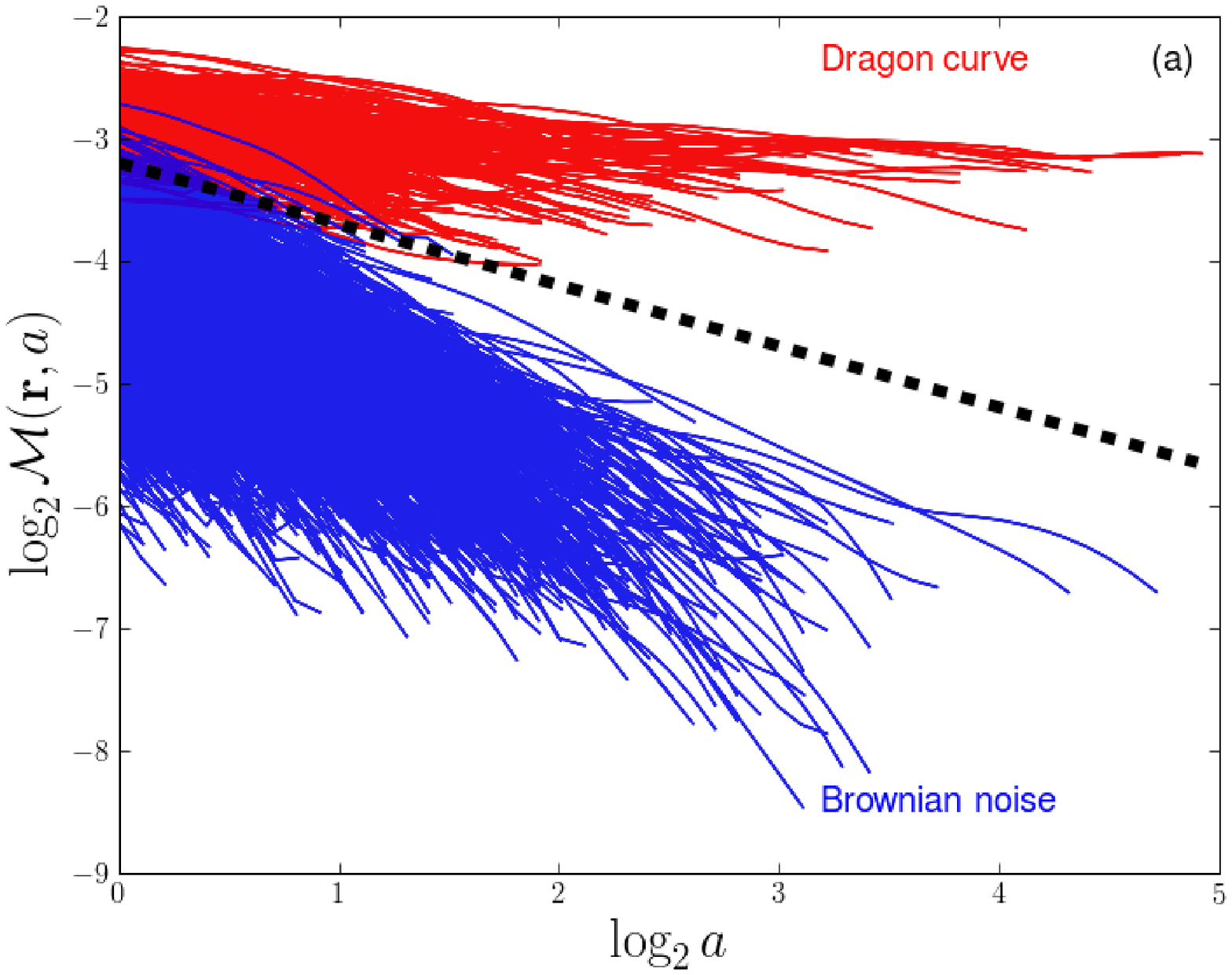}{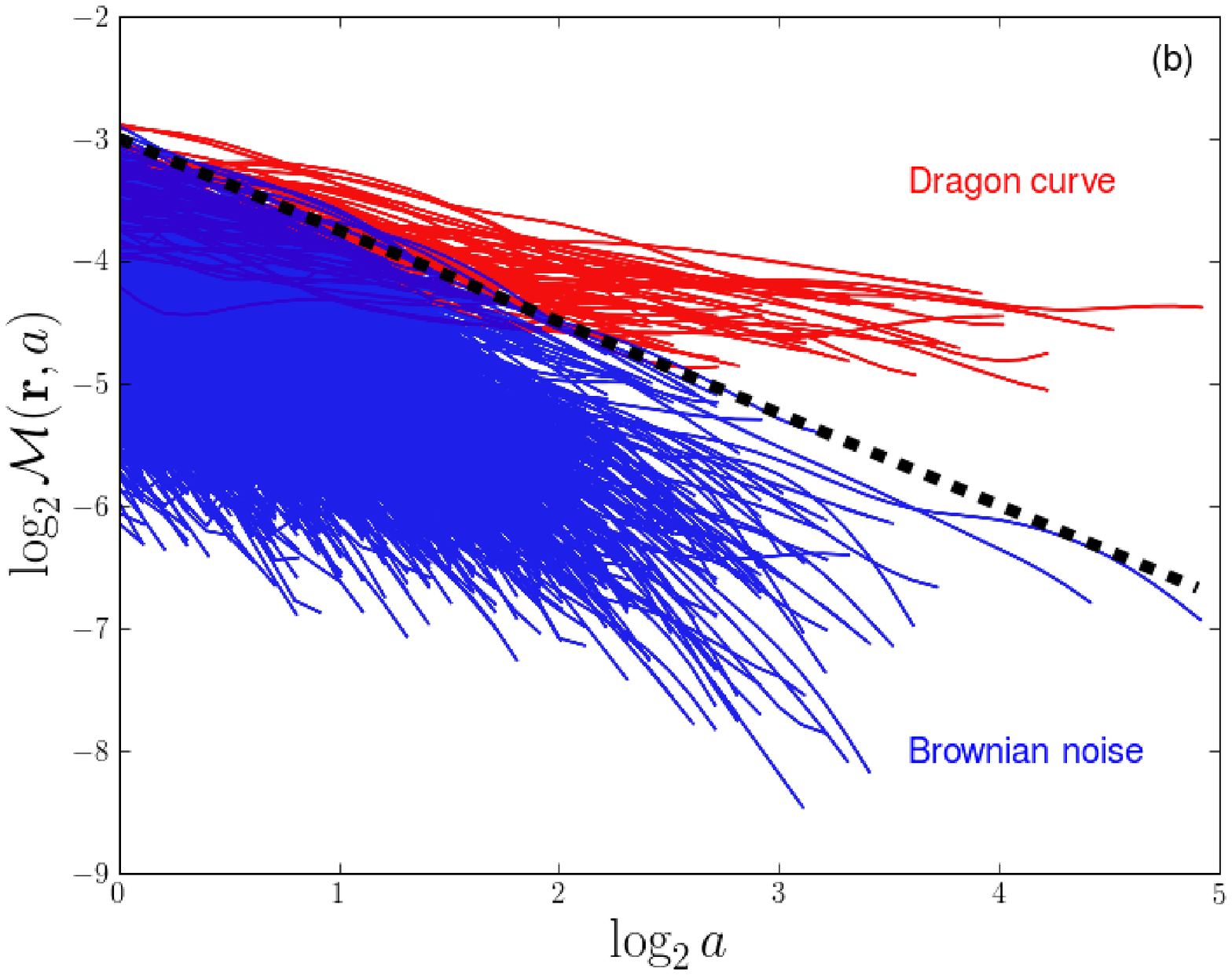}
\caption{Log-log plot of \WT{} modulus along the skeleton maxima
  lines versus scale. Lines are colored according
  to the segmentation procedure (Eq.~(\ref{eq_wtmm_12})) : fractal
  Dragon boundary (red) and fractionnal Brownian noise (blue). (a)
  Noisy fractal Dragon with a noise amplitude twice as large as the
  fractal Dragon (see Fig.~\ref{fig_dragon}(d)).
 (b) Noisy
  fractal Dragon with a noise amplitude five times as large as the
  fractal Dragon. The dashed black line represents the segmentation condition (Eq.~\ref{eq_wtmm_12}).}
\label{fig_gerbe_dragonBro}
\end{figure}
As previously  emphasized~\citep{aMuz94,aArn95,bArn02}, the set of
maxima lines that defines the \WT{} skeleton contains the
space-scale information necessary to
recover the underlying multifractal properties.
In Figures~\ref{fig_gerbe_dragonBro}(a) and
\ref{fig_gerbe_dragonBro}(b) are shown in a logarithmic
representation, the behavior of the 
\WT{} modulus along the maxima lines computed for a noisy fractal
Dragon with a noise amplitude respectively twice and five times as
large as the fractal Dragon. Since the fractional Brownian noise is
everywhere singular with H\"older exponent $H=-0.7$~\citep{bMan82},
maxima lines pointing to noise features at small scale are
characterized by a \WTMMM{} power law behavior $\Mpsi[f](a)
\sim a^{-0.7}$, while lines associated to the fractal Dragon boundary
can be distinguished by the fact that $\Mpsi[f](a) \sim a^{0} \sim
Const$ (no scale dependence). This leads us to implement the following
segmentation procedure: the space $(\log_2 a, \log_2 \Mpsi[f](a))$
is divided in two regions separated by a straight line of slope
$-0.7<h_s<0$ and intercept $\log_2 M_s$. As shown in
Figure~\ref{fig_gerbe_dragonBro}(a), for low noise amplitude, all the
maxima lines along which $\log_2 \Mpsi[f](a)$ decays slower than $h_s
\log_2 a$ when increasing $a$, are colored in red and associated with
the Dragon boundary.
On the contrary, all the maxima lines along which $\log_2 \Mpsi[f](a)$
decays faster than $h_s \log_2 a$ when increasing $a$, are colored in
blue and associated with the noise component. But as shown in
Figure~\ref{fig_gerbe_dragonBro}(b), for large noise amplitude the
distinction of the two sub-skeletons is much more tricky at small
scales where some entangling is observed. We thus adapt the
segmentation criteria towards the largest scales in fully analogy with
a different but conceptually similar adaptation of the \DD{} \WTMM{}
segmentation method~\citep{kha07}.
Each maxima line is characterized by a length, \textit{i.e.} its maximun scale $a_{max}$ and the
\WT{} modulus $\Mpsi[f](a_{max})$ at that scale. A given maxima
line is said to belong to the Dragon sub-skeleton, if it satisfies the
following condition :
\begin{equation}
  \log_2 \Mpsi[f](a_{max}) \ge h_s \log_2 a_{max} + \log_2 M_s.
\label{eq_wtmm_12}
\end{equation}

\begin{figure}
\centerline{\includegraphics[width=0.65\linewidth,angle=-90]{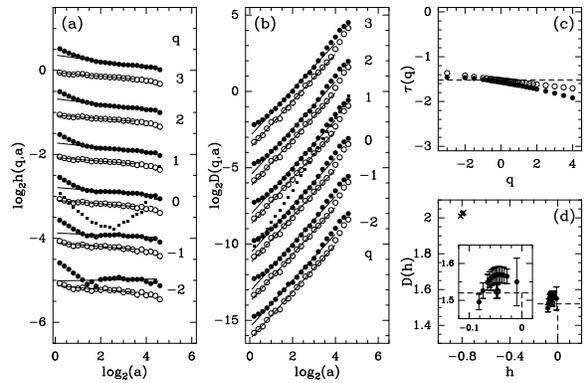}}
\caption{Multifractal analysis of the fractal Dragon
 ($\whiteCircle$) and of the noisy ($H=-0.7$) fractal Dragon
 ($\blackCircle$) after applying the
 segmentation procedure (Eq.~(\ref{eq_wtmm_12})). (a) $h(q,a)$ vs
 $\log_2 a$ for different values of $q$; the solid lines correspond to
 linear regression fits over the range of scales $a \in [2^0,2^4]~\sigma_W$
 (resp. $[2^{0.5}, 2^{3.5}]~\sigma_W$) for the fractal Dragon (resp. the noisy fractal Dragon
 after segmentation). The symbols ($\blackSquare$) correspond to the $h(q=0,
 a)$ partition function obtained for the noisy fractal Dragon without any
 segmentation. (b) $D(q,a)$ vs $\log_2 a$. (c) $\tau(q)$ vs $q$; the
 dashed horizontal line is the theoretical spectrum $\tau(q)=-1.5236$
 ($\forall q$) of the
 boundary of the fractal Dragon. (d) $D(h)$ vs $h$; the symbols
 ($\blackCircle$) correspond to the segmented $D(h)$ spectrum for
 the fractal Dragon component; the ($\times$) correspond to
 the extracted $D(h)$ spectrum of the colored noisy background. The
 zoom-in window enlarges $D(h)$ data corresponding to the noisy fractal Dragon
 ($\blackCircle$) after applying the
 segmentation procedure. }
\label{fig_pf}
\end{figure}
\begin{figure}
\plottwo{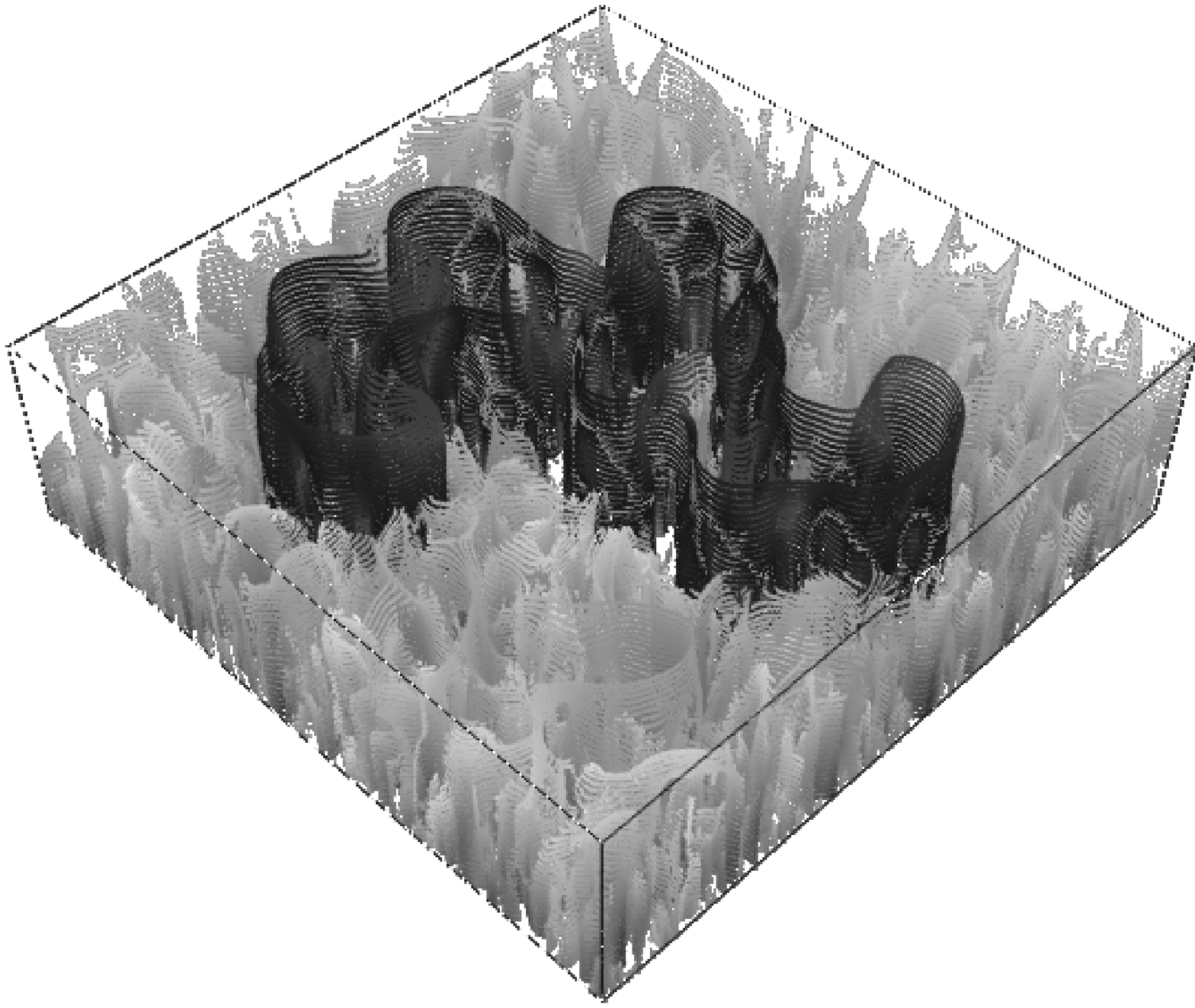}{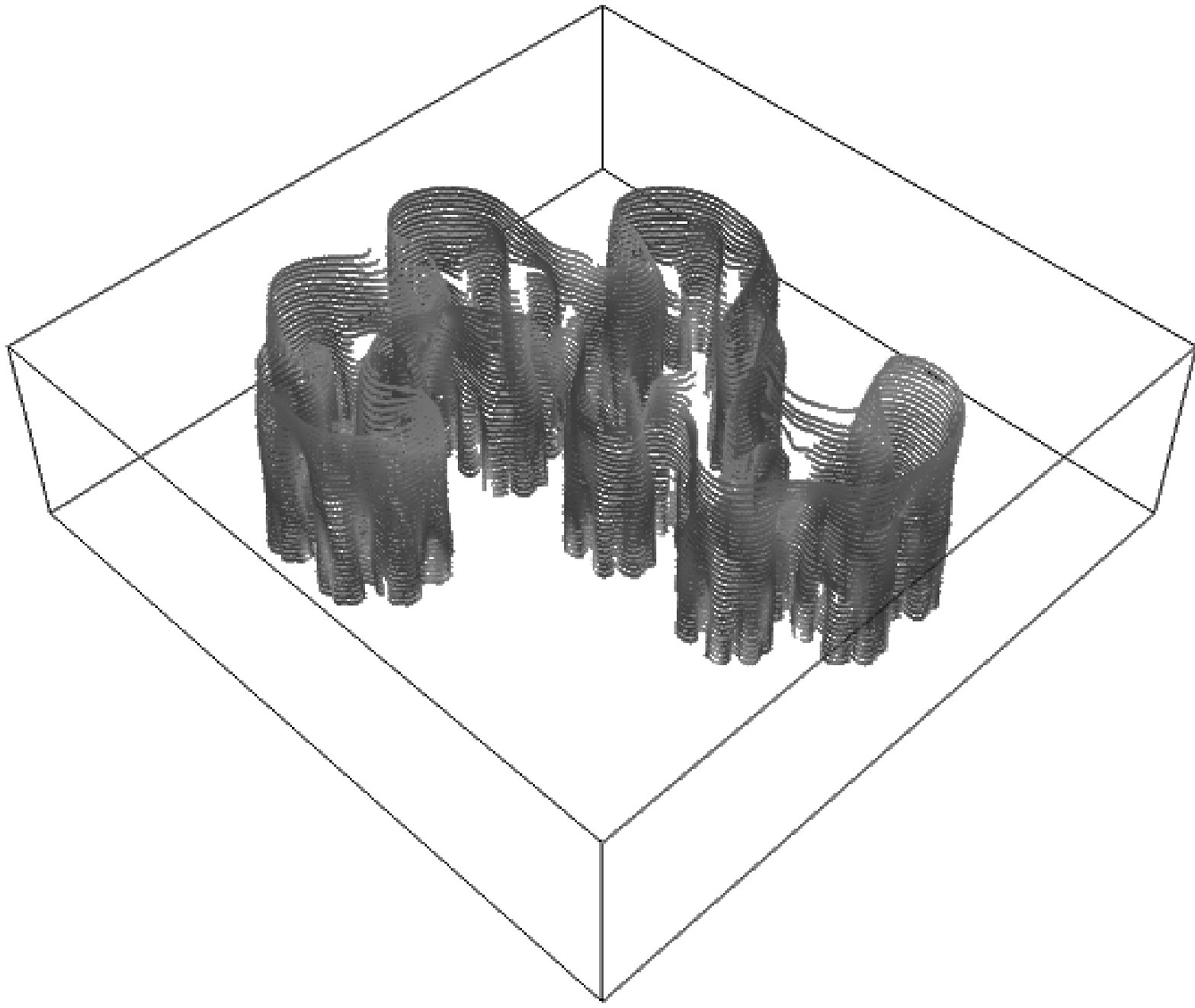}
\caption{3D visualization in the space-scale $(x,y,scale)$ representation of the \WTMM{}
 chains computed from the image shown in Fig.~\ref{fig_dragon}(d)
 before (a) and after (b)
 the segmentation procedure (Eq.~(\ref{eq_wtmm_12})).
At each scale $a$, only the maxima chains containing at
least one \WTMMM{} belonging to the resulting \WT{} skeleton are displayed.}
\label{fig_skeleton}
\end{figure}

In Figures \ref{fig_pf}(a) and \ref{fig_pf}(b) are reported the results of the
computation of the partition functions for the fractal Dragon alone and
its noisy version after applying the segmentation condition (Eq.~(\ref{eq_wtmm_12}))
with $h_s = -0.5$ and $\log_2 M_s = -3.2$. 
In Figure \ref{fig_pf}(a), the partition functions $h(q,a)$
(Eq.~(\ref{eq_wtmm_5}))
 of the
noisy Dragon display
a well defined scaling behavior over 3 octaves (compared
to 4 octaves for the original fractal Dragon), for a wide
range of values of $q \in [-2, 3]$. 
For negative $q$ values, at very small scales, the segmentation procedure
fails to disentangle the two components due to the contribution from
the noisy component with $h(q)\simeq-0.7$. Nevertheless the gain in scaling is
unquestionable as compared to the behavior of the $h(q,a)$ partition
functions without segmentation (the $\blackSquare$ in Fig.~\ref{fig_pf}(a)).
Without any segmentation the partition functions are a mixture of different
scaling behaviors from which reliable quantitative information cannot
be extracted. In Figures \ref{fig_pf}(c) and \ref{fig_pf}(d) are shown
the corresponding $\tau(q)$ and $D(h)$ spectra. Despite some slight
departure from monofractality for the segmented noisy fractal Dragon
(that is also observed but to a lesser extent in the original fractal
Dragon as the result of finite-size effects), one recovers a rather
good estimate of the fractal dimension $D_F=1.57 \pm 0.03$ of the
fractal Dragon boundary. Furthermore, as reported in Figure
\ref{fig_pf}(d), our segmentation procedure has proved to be very
efficient to estimate separately the $D(h)$ singularity spectra of
both the fractal Dragon and the noisy background. This efficiency is
illustrated in Figure \ref{fig_skeleton}
where a \TD{} $(x,y,scale)$ space-scale visualization of the maxima chains of
the noisy Dragon prior (Fig.~\ref{fig_skeleton}(a)) and after (Fig.~\ref{fig_skeleton}(b)) segmentation clearly confirms
the elimination of noise-induced small scale features that would otherwise
severely affect the multifractal analysis.

%
\section{Application of the wavelet-based segmentation method to Solar
  magnetogram data}
\label{observations}
Magnetic field measurements were obtained by the Michelson Doppler Imager (MDI) on the \emph{Solar and Heliospheric Observatory} (SOHO), which images the Sun on a 1024$\times$1024 pixel CCD camera through a series of increasingly narrow filters~\citep{1995SoPh..162..129S}. The final elements, a pair of tunable Michelson interferometers, enable MDI to record filtergrams with an FWHM bandwidth of 94~m\AA. In this paper, 96-minute magnetograms of the full disc were used, which had a pixel size of $\sim$2". 
For the purposes of this work, a series of magnetograms have been
analyzed to examine the difference in fractal properties between quiet
and active solar regions. A total of 29 magnetograms representative of
the quiet Sun were taken from December 21 to December 22, 2006 and a
similar series of 28 images representative of the active Sun were
taken from October 27 to October 29, 2003.
In the solar photosphere, the large magnetic Reynolds number ($\sim
10^7 - 10^9$) means that magnetic field lines will be advected with the
flow of plasma \citep{McAteer2009}. This system naturally leads to
self-similarity, suggesting a multifractal study is appropriate
\citep{Lawrence1993-apj,Abramenko2002-apj,McAteer2005-apj}.
As already mentioned, previous method of calculating the multifractal properties of solar magnetic features are dependent on image resolution, thresholding, and instrument sensitivity. The \WTMM{} method calculates the multifractal spectrum of solar magnetic features based on the distribution of gradients within the image at various scales. As such, the \WTMM{} multifractal parameters are less sensitive to changes in image resolution and instruments than traditional methods.

\subsection{Quiet-Sun multifractal properties}
\label{sect_QS}

\begin{figure}
\centerline{\includegraphics[width=0.85\linewidth]{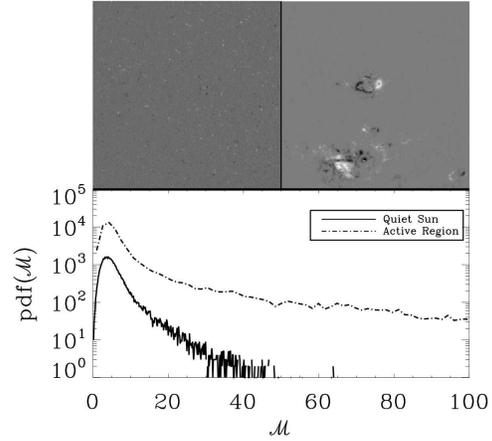}}
\caption{(Top, Left) MDI magnetogram taken on December 20, 2006; (Top,
  Right) MDI magnetogram taken on October 28, 2003. (Bottom) Histogram
  values of the wavelet transform modulus $\Mpsi[f](a)$ at the
  smallest scale ($\sigma_W=7$~pixels) for MDI magnetogram images of a quiet Sun (solid) and active Sun (dashed).}
\label{fig_qs}
\end{figure}
Examples of quiet and active MDI magnetograms analyzed are shown
in Figure \ref{fig_qs} (top left and top right
respectively), with a histogram of the wavelet 
transform modulus $\Mpsi[f]({\bf b}, a)$ at the smallest scale
(bottom). Active regions result from an increased
proportion of large magnetic elements of opposite polarity in close
proximity to each other. The resulting neutral or magnetic inversion
lines can be detected using standard wavelet-based
techniques~\citep{Ireland:2008}. As such active regions should contain
a greater number of higher magnitude \WT{} gradients. This is shown in
Figure \ref{fig_qs} (bottom), which suggests that moduli with
values larger than $40$ are unlikely in quiet-Sun magnetograms. Due
to the different scaling properties of active regions and their
surrounding quiet Sun, our goal is to segment the \WT{} skeletons
using the condition defined in Eq.~(\ref{eq_wtmm_12}). 
As outlined in Section~\ref{segmentation} and illustrated on synthetic
data in Section \ref{section_synthetic_data}, this should allow us to
study the multifractal properties of active regions in a
quantitative manner.

\begin{figure}
\centerline{%
\includegraphics[angle=-90,width=0.99\linewidth]{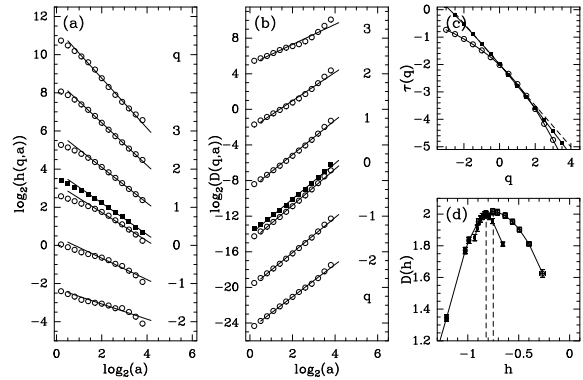}}
\caption{Multifractal analysis of a set of 30 quiet-Sun images (505$\times$505)
 prior ($\whiteCircle$) and after ($\blackSquare$) thresholding (a
 sample thresholded magnetogram is shown in Figure \ref{fig_qs_image}).
 (a) $\log_2 h(q,a)$ vs $\log_2 a$ for different values
 of $q$; the solid lines are
 linear regression fits over the range of scales $a \in [2^0,2^{3.7}]~\sigma_W$.
 (b) $\log_2 D(q,a)$ vs $\log_2 a$. 
 (c) $\tau(q)$ vs $q$; the
 dashed straight line is the theoretical linear spectrum $\tau(q)=-0.75q-2$ of the
 \DD fractional Brownian noise with H\"older exponent $H=-0.75$. 
 (d) $D(h)$ vs $h$; the dashed lines delimit the position of the top
 of the $D(h)$ curves. Error bars correspond to standard deviation in
 the linear regression procedure.}
\label{fig_qs_pf}
\end{figure}
\begin{figure}
\centerline{%
\includegraphics[angle=90,width=0.45\linewidth]{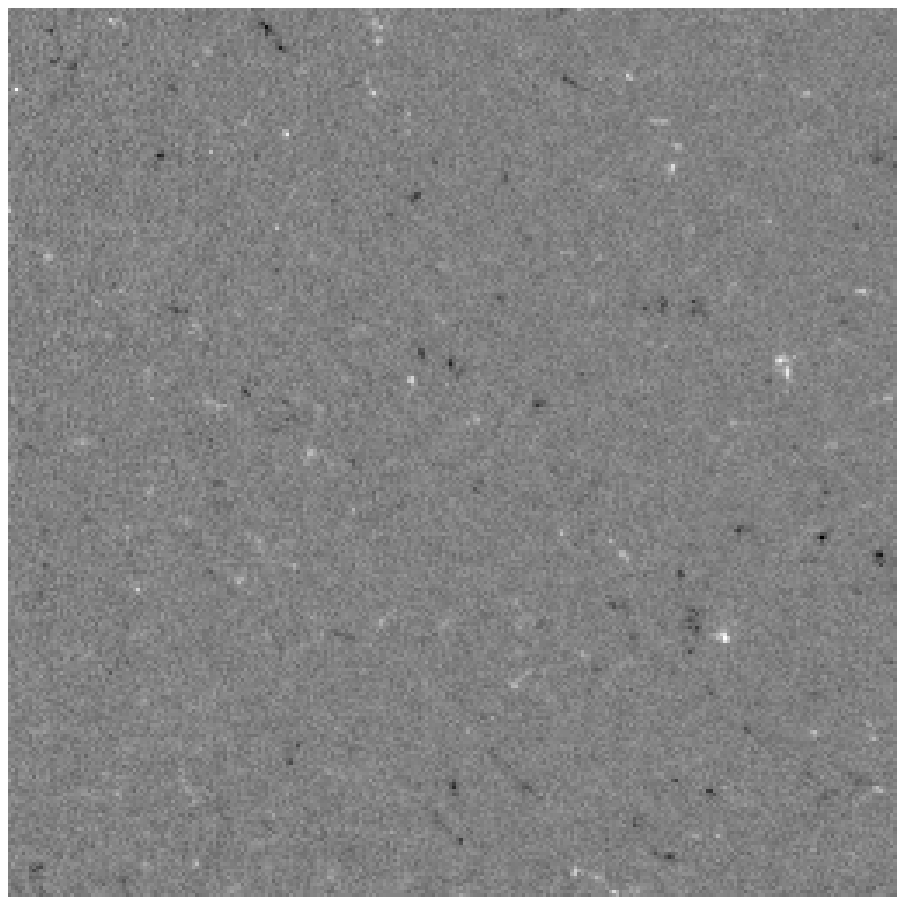}
\includegraphics[angle=90,width=0.45\linewidth]{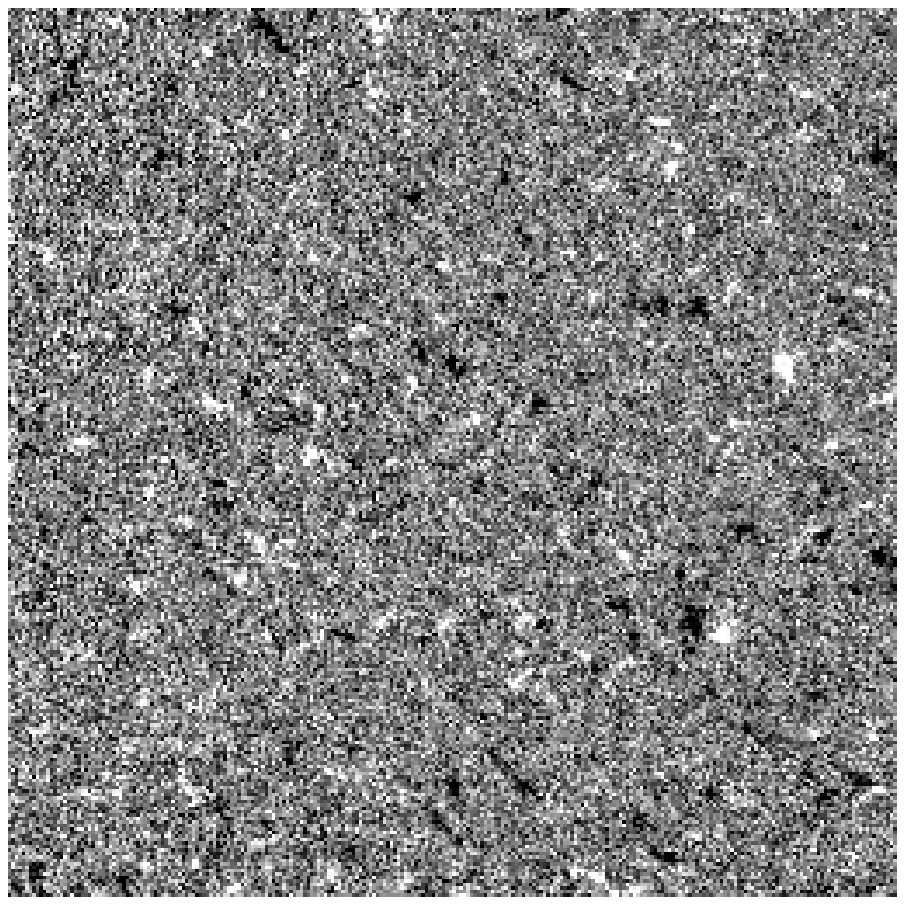}}
\caption{256$\times$256 quiet-Sun images. Image on the right is a
 threshholded version of the left one. Pixels with large absolute
 magnetic flux are shrinked down. Multifractal properties of quiet-Sun
 images and the corresponding thresholded versions are shown in
 Figure \ref{fig_qs_pf}.}
\label{fig_qs_image}
\end{figure}
The results of the computation of the multifractal spectra when 
averaging the partition functions over a set of 30 
($505 \times 505$) quiet-Sun images without applying the
segmentation are reported in Figure~\ref{fig_qs_pf}. 
As shown in Figure~\ref{fig_qs_pf}(a) and \ref{fig_qs_pf}(b), $h(q,a)$
(Eq.~(\ref{eq_wtmm_5})) and $D(q,a)$ (Eq.~(\ref{eq_wtmm_6})) 
display convincing scaling behavior over almost four octaves
 for $q \in [-2, 3]$ (symbol ($\whiteCircle$)). 
Linear regression fits of the data yield 
the non-linear $\tau(q)$ spectrum shown in Figure \ref{fig_qs_pf}(c). 
This multifractal diagnosis can also be observed in Figure 
\ref{fig_qs_pf}(a) where the slope $h(q)$ of the partition function 
$h(q,a)$ versus $\log_2 a$ definitely depends on $q$. The corresponding 
multifractal spectrum $D(h)$ is shown in Figure \ref{fig_qs_pf}(d). 
From the top of the $D(h)$ curve, we can 
see that quiet-Sun images are everywhere singular ($D(q=0)=2$) 
with a corresponding H\"older exponent $h(q=0) \simeq -0.75$. 
The multifractality can be quantified by the so-called intermittency
coefficient $c_2$ that characterizes the width of the $D(h)$ curve. As
shown in Figures \ref{fig_qs_pf}(c) and \ref{fig_qs_pf}(d), the
$\tau(q)$ and $D(h)$ data of the quiet-Sun images are well fitted by a parabola
\begin{equation}
\tau(q) = -c_0+c_1q-\frac{c_2}{2}q^2, \;\;\;\;\;\; D(h) = c_0  - \frac{(h-c_1)^2}{2 c_2},
\label{eq_wtmm_13}
\end{equation}
where 
$c_0 \simeq 2$, $c_1 \simeq -0.75$ and $c_2 \simeq 0.22$.
Let us point out that quadratic multifractal spectra are predicted by
the so-called log-normal model that has been popularized by the
fully-developed turbulence
community~\citep{bFri95,aArn98,aArn99,aDelArn01}. In the present case,
there is no particular evidence of the relevance of this model except
that the observed $\tau(q)$ and $D(h)$ multifractal spectra are well
characterized by their log-normal quadratic approximations.

In order to understand the source of this intermittency, a upper threshold
was imposed on each MDI magnetogram of the quiet Sun
(Figure~\ref{fig_qs_image}). The threshold operation has the effect of
removing large magnetic features resting on the boundary of the super-granular
structures of the Sun. In Figure \ref{fig_qs_pf}(d), we can see that
the multifractality of the thresholded quiet-Sun image (symbol
$\blackSquare$) set is strongly
reduced but not totally cancelled. We can also note that the average
H\"older exponent is slightly shifted from $<h>=c_1=-0.75$ to $-0.82$, and
the intermittency coefficient is reduced to $c_2 \sim 0.06$ (Let us
recall that a value $c_2=0$ means that the underlying process is monofractal). Without
the super-granular magnetic structure, the quiet-Sun multifractal
spectrum looks much more monofractal. This suggests that the magnetic
features resting on the boundaries of the super-granular structure are a major actor in
the observed intermittent structural properties of the Sun~\citep{Georgoulis2002-apj}. Since
current models for the solar dynamo use information on the fractal
dimension of solar disk as a whole~\citep{Pontieri:2003}, these new
informations on the photosphere and the characteristic make-up of the
quiet Sun should be incorporated in further theoretical works.

\subsection{Solar magnetogram active region segmentation}
\label{sec:sol_mag_active}
In this section, we highlight the use of the \WTMM{} segmentation
method on Solar magnetogram data with active regions, to demonstrate
its ability to analyze the underlying multifractal properties of the active
regions that are embedded in the surrounding quiet-Sun texture.

\begin{figure}
\centerline{%
\includegraphics[angle=-90,width=1\linewidth]{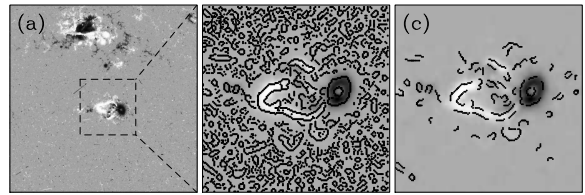}%
}
\caption{(a) 505$\times$505 Active Region example image (October 28,
  2003). (b) \WTMM{} chains at the smallest scale ($a=\sigma_W$~pixels)
 in a small $150\times 150$ region surrounding an active spot. (c)
 \WTMM{} chains left after the segmentation step.}
\label{fig_ar_image}
\end{figure}
\begin{figure}
\centerline{%
\includegraphics[width=0.99\linewidth]{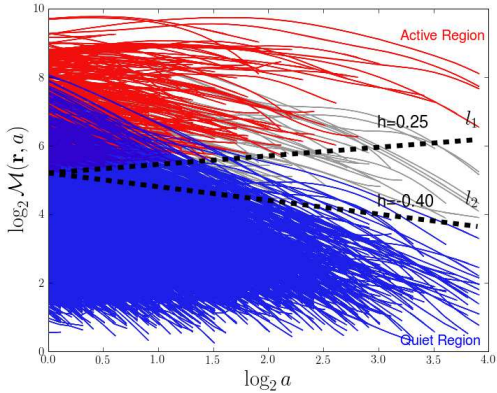}
}
\caption{Log-log plot of the \WT{} modulus along the skeleton
  maxima lines versus scale. The dashed line
denoted $l_1$ with slope $h_A=0.25$ and intercept
$\log_2 M_A = 5.2$ is used to identify
\WT{} skeleton maxima lines associated to the active region
(red). According to Eq.~(\ref{eq_wtmm_15}), these lines have an ending point at highest scale ($\log_2 a_{max},
\log_2 \Mpsi[f](\mathbf{r},a_{max})$) above $l_1$.
The dashed line denoted $l_2$ with slope $h_Q=-0.40$ and intercept
$\log_2 M_Q = 5.2$ is used to identify
maxima lines associated to the quiet Sun (blue). According to
Eq.~(\ref{eq_wtmm_14}), these lines
have an ending point below $l_2$. All other lines (grey) are not clearly
identified to belong either to the active site or the quiet
surrounding. The values of the segmentation parameters $h_A$, $\log_2
M_A$, $h_Q$ and $\log_2 M_Q$ were chosen by examining the \WTMM{}
histogram at the smallest scale to extract at best a sub-skeleton
specific to the active region.}
\label{fig_skeleton_ar}
\end{figure}
\begin{figure}
\includegraphics[width=\linewidth]{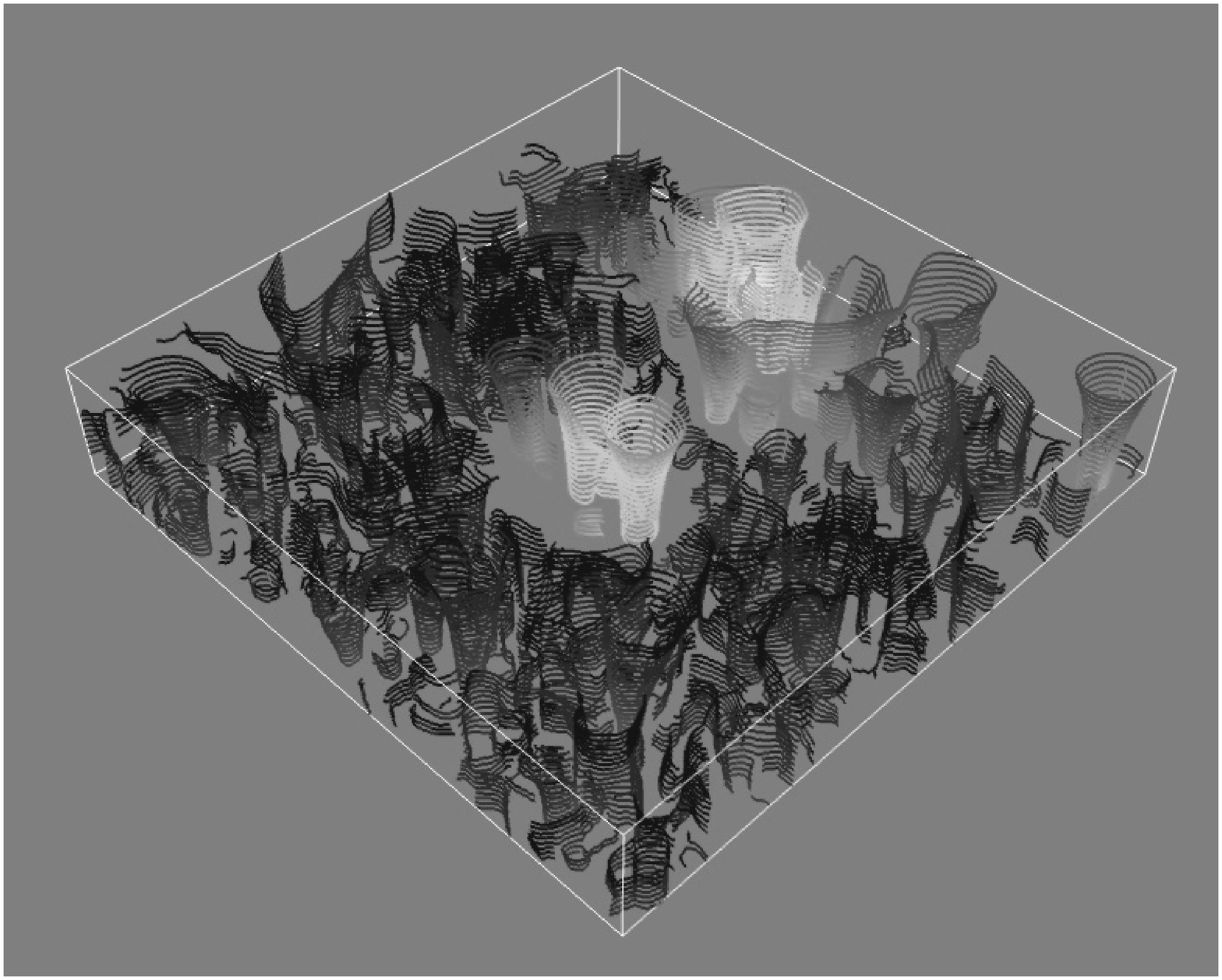}\\
\includegraphics[width=\linewidth]{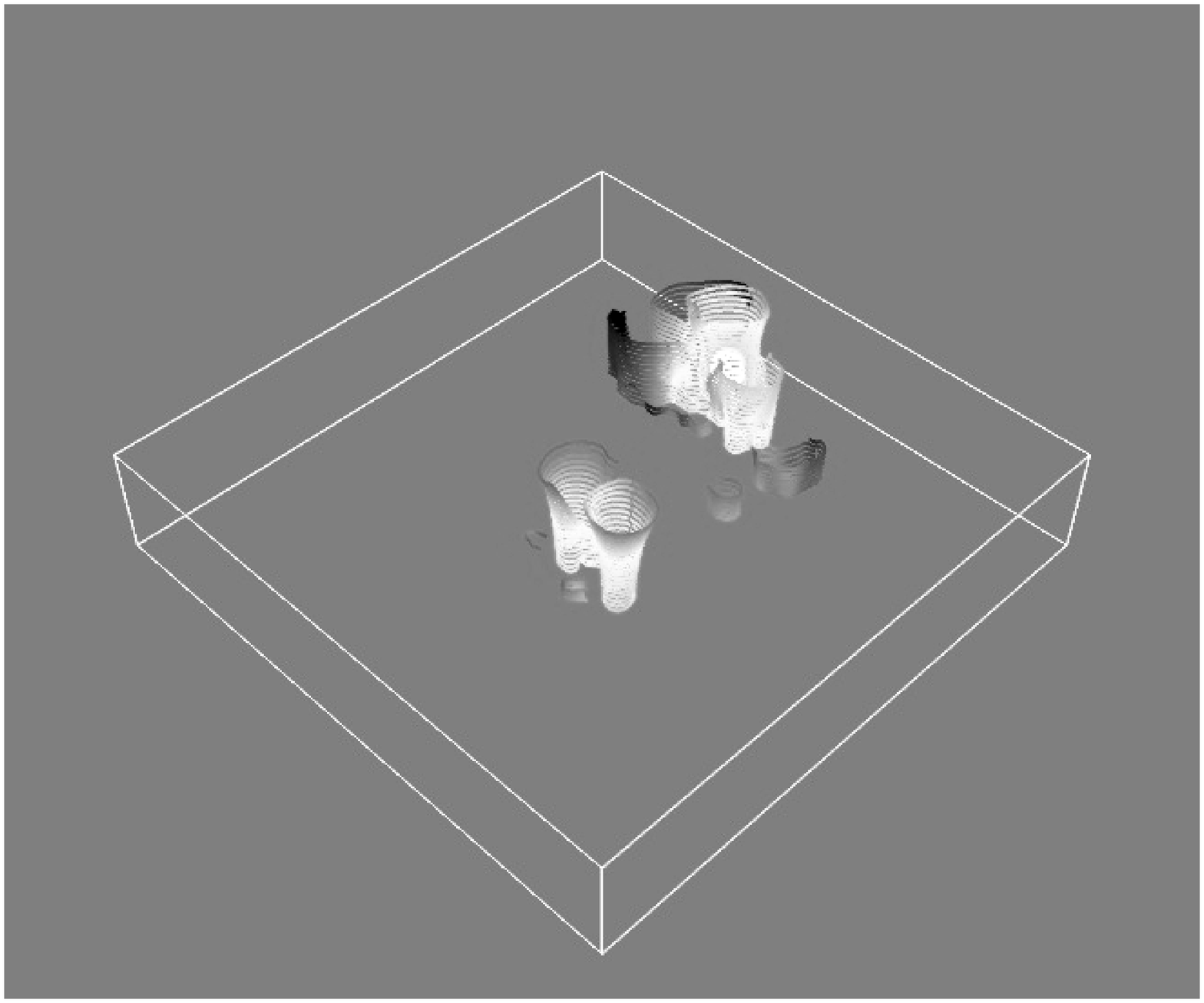}
\caption{3D visualization in the space-scale $(x,y,scale)$ representation of the
 \WTMM{} 
 chains computed from the image shown in Fig.~\ref{fig_ar_image}(a)
 before (left) and after (right)
 the segmentation procedure. The segmentation conditions are defined
 in Figure~\ref{fig_skeleton_ar}. The maxima
 chains displayed contain at least one \WTMMM{} belonging to the resulting skeleton.}
\label{fig_skeleton_ar2}
\end{figure}

A sample $505\times 505$
magnetogram MDI image containing an active
region is shown in
Figure~\ref{fig_ar_image}. Figures~\ref{fig_ar_image}(b) and
\ref{fig_ar_image}(c) show respectively the 
results of the computation of the \WTMM{} chains before and after the
segmentation at scale $a=\sigma_W \sim 7$ pixels of a small $150\times 150$ excerpt focused on the active location. 
As explained in Section~\ref{sect_QS}, \WT{} skeleton maxima lines
associated with quiet-Sun structures have a characteristic scaling behavior
described by $\Mpsi[f](a) \sim a^{-0.7}$. This behavior is used to
derive the parameters characterizing the 
line $l_2$ in
Figure~\ref{fig_skeleton_ar}
that will allow us to discriminate in the \WT{} skeleton, the maxima
lines (blue) that correspond to quiet-Sun structures:
\begin{equation}
  \log_2 \Mpsi[f](a_{max}) \le h_Q \log_2 a_{max} + \log_2 M_Q,
\label{eq_wtmm_14}
\end{equation}
where $-0.7<h_Q<0$, so that for the selected maxima lines $\log_2
\Mpsi[f](a)$ will decrease fast enough when increasing the scale $a$
to correspond to the quiet phase. To select the maxima lines (red)
associated with the active region, we use another separating line
$l_1$ in Figure~\ref{fig_skeleton_ar}:
\begin{equation}
  \log_2 \Mpsi[f](a_{max}) \ge h_A \log_2 a_{max} + \log_2 M_A,
\label{eq_wtmm_15}
\end{equation}
where this time $0\le h_A < 0.38$, to limit the decrease (if any) of
$\log_2 \Mpsi[f]$ when increasing $a$. Note that
the lines  $l_1$ and $l_2$ have different slopes
because some maxima lines cannot be clearly associated with either the
quiet Sun or the active region state. Indeed those maxima lines are
associated to features located near the boundary between quiet-Sun and active regions.
When going from small scales to
large scales the support of the analyzing wavelet starts covering partly both
regions preventing accurate classification. 
As illustrated in Figure~\ref{fig_skeleton_ar2}, when fixing the
segmentation parameters to $h_Q=-0.40$, $h_A=0.25$ and $\log_2 M_Q =
\log_2 M_A = 5.2$, 
the space-scale nature of the methodology allows us to disentangle maxima
chains associated with the active region from those corresponding
to the quiet Sun. Let us note that in a future work on large data sets, an
automated parameters adjustment will be implemented using a clustering algorithm.
In addition, a wrong choice of segmentation parameters can be observed in the
partition function plots (not shown here) which display a phase
transition phenomenon,
i.e. a scaling behavior that changes from one state to the other when
going from small scales to large scale due to to non-homogenous phases
and mis-classified maxima lines.
%
%
%
\begin{figure}
\includegraphics[angle=-90,width=1.0\linewidth]{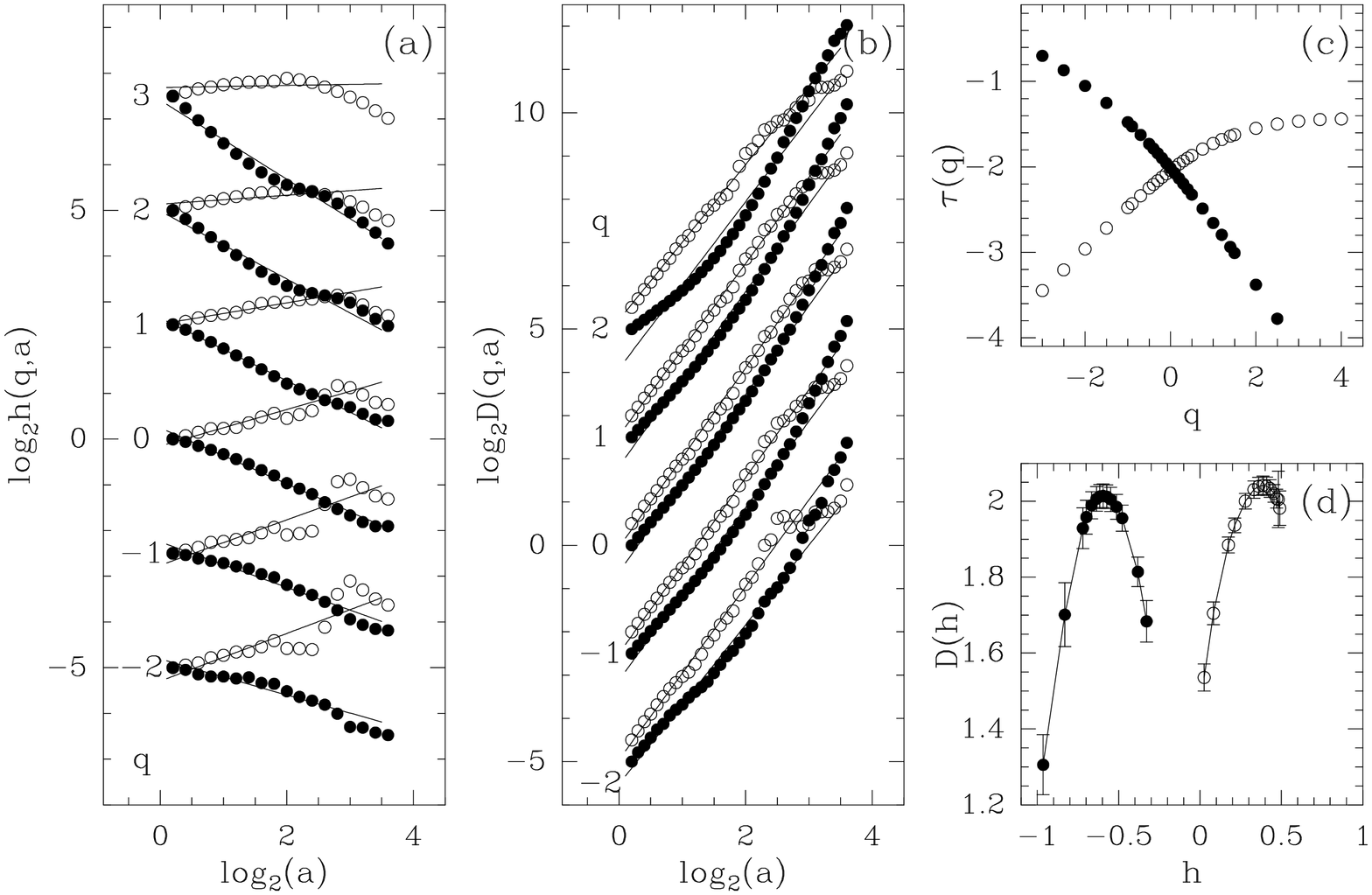}
\caption{Multifractal analysis of a set of 5 active region magnetogram
  images (505$\times$505) corresponding to the two sub-skeletons
  identified in Figures \ref{fig_skeleton_ar} and
  \ref{fig_skeleton_ar2}. The symbols ($\blackCircle$) correspond to the segmented quiet Sun
  and ($\whiteCircle$) to the segmented active region.
 (a) $h(q,a)$ vs $\log_2 a$ for different values
 of $q$; the solid lines are
 linear regression fits over the range of scales $a \in [2^0,2^{3.0}]~\sigma_W$.
 (b) $D(q,a)$ vs $\log_2 a$. 
 (c) $\tau(q)$ vs $q$.
 (d) $D(h)$ vs $h$. Error bars correspond to standard deviations in the
 linear regression procedure.
}
\label{fig_ar_pf}
\end{figure}

In Figure~\ref{fig_ar_pf} are shown the results of the
partition function computation for a set of 5 active region
magnetogram images taken on October 28$^{th}$, 2003 (one image out of
this set is shown in Figure~\ref{fig_ar_image}(a)). 
Partition functions are computed separately for each sub-skeleton
corresponding to the extracted action region maxima lines and
quiet-Sun maxima lines shown in Figure~\ref{fig_skeleton_ar}.
From these plots, one can see that the $\log_2 h(q,a)$ versus $\log_2
a$ data are nicely modeled with linear regression fits with slope
$h(q)$ that depends on $q$ as the signature of multifractal
scaling. This demonstrates that the segmentation procedure
successfully extracts from the magnetogram images two scale invariant
components with different multifractal properties. The $\tau(q)$
(Fig.~\ref{fig_ar_pf}(c)) and $D(h)$ (Fig.~\ref{fig_ar_pf}(d))
spectra computed from the set of quiet-Sun maxima lines (blue lines in
Fig.~\ref{fig_skeleton_ar}) are in good agreement with the ones
previously computed in Section \ref{sect_QS} from pure quiet-Sun
images (Figs. \ref{fig_qs_pf}(c) and \ref{fig_qs_pf}(d) respectively).
Using
the log-normal approximation (Eq.~(\ref{eq_wtmm_13})), we get 
$c_0=2$, $c_1 = -0.65$ and $c_2 = 0.10$. This means
that the extracted quiet Sun appears (like the thresholded MDI
magnetograms of quiet Sun in Figs \ref{fig_qs_pf}(c) and \ref{fig_qs_pf}(d))
a little less intermittent as compared
to the previous estimate $c_2=0.22$.
As for the active region, the corresponding partition functions
computed from the set of active phase maxima lines (red lines in
Fig.~\ref{fig_skeleton_ar}) display very convincing multifractal
scaling behavior as quantified by the $\tau(q)$ and $D(h)$ spectra
shown in Figures \ref{fig_ar_pf}(c) and \ref{fig_ar_pf}(d)
respectively. Again these spectra are well approximated by the
quadratic log-normal formula (Eq.~(\ref{eq_wtmm_13})) with parameters
$c_0=2$, $c_1 = 0.38$ and $c_2 = 0.12$. This indicates that the
singularities associated with the active region are space-filling
(they are distributed on a set of fractal dimension $D_F=c_0=2$), with
a mean strength $h(q=0)=c_1=0.38$ meaning that magnetogram images can
be considered as continuous on active regions (over the range of scales of our analysis) but noisy on quiet regions. 


%
\section{Conclusions}
\label{conclusions}

Many complex physical systems analyzed using fractal and multifractal
techniques are surrounded or embedded in a noisy background sometimes
originating from instrumental noise. As such these systems are a
statistical combination of two distinct self-similar structures. This
work addressed the need for an accurate calculation of the multifractal
parameters of such complex systems. The presence of compound
scale-invariant structures can result in an inaccurate or skewed
calculation of the fractal and multifractal parameters when studied as
a whole. 
Using a wavelet-based multi-scale segmentation method, we show that it
is possible to disentangle to some extent these two processes and
accurately (up to finite-size effects) recover the multifractal
characteristics of the system of interest. A theoretical test example
for this method was provided in section \ref{segmentation}. The
removal of information relating to the background noise was
highlighted in Figure~\ref{fig_skeleton}. The quantitative results
reported in Figure~\ref{fig_pf} attest of the ability of this
segmentation method to recover the multifractal parameters in question.

Let us emphasize here that the application of this method to
experimental data for which we do not have a priori knowledge of the
possible underlying multifractal processes is a difficult task that
requires much attention to perform the most objective segmentation
which can not be infered or guided by some physical rule or
information. The multifractal analysis is a statistical tool that has
direct connections with signal and image processing, but not
necessarily with the physics of the system per se.
As noticed in section \ref{sec:sol_mag_active}, the use of a clustering
algorithm should greatly help in adjusting the multifractal parameters
of the differents components as well as in providing an automated
procedure for processing large data sets. This will be reported in a
future publication.

The application of this wavelet-based methodology to quiet-Sun data
has revealed the multifractal nature of this intermittent noisy
component ($<h>=-0.75$) as mainly resulting from the super-granular
magnetic structures. The quiet-Sun study was also necessary to get
expertise for further analyzing more complex images that involve a
segmentation before being able to clearly identify the underlying
multifractal properties. We have checked
that the partition functions computations for the segmented quiet-Sun
phase provide (i) convincing scaling properties and (ii) multifractal
spectra $\tau(q)$ and $D(h)$ estimates 
in good numerical agreement with the one measured in the previous calibration
step.  The assumption of two non-overlapping $D(h)$ is not
inconsistent with the data. Let us notice that a phase
transition can be observed in the partition function log-log plots
when inappropriate segmentation parameters are chosen. In the case
of overlapping $D(h)$, there would be a large set of maxima lines in
the WT skeleton that could not be genuinely sorted, which would
prevent from building accurate multifractal $D(h)$ measurement.
From the analyzed data, we were not able to distinguish more than two
phases. Finally, this gives us good confidence in the segmentation proposed for solar
magnetogram containing an active region. However, further study is
needed to precisely quantify scaling properties associated to specific
active region features (\textit{e.g.} emerging magnetic flux along the main
 polarity inversion line, sunspots build-up, delta-configuration...)
 and how the \WTMM{} method can be sensitive to these elements. More
 precisely, when analyzing higher resolution images than MDI, we
 expect that this segmentation tool will be all the more necessary as
 quiet-Sun and active region features are more entangled.

The main outcome of
the present study is the demonstration that the proposed multi-scale
segmentation procedure provides an objective way of studying the
complexity in active regions separately from the surrounding quiet
Sun. As such our results are significantly more stable and robust when
compared to previous fractal and multifractal
analysis~\citep{Pontieri:2003,McAteer2005-apj}. In a forthcoming paper, we
will report on the application of this segmentation method to
characterize the evolution of active regions keeping
track of the multifractal parameters for possible correlations with
extreme solar events~\citep{Gallagher:2007}. Other applications of this method are
in progress as the analysis of the intrinsic multifractal
properties of entangled hot and cold interstellar atomic gas from
\TD{} numerical simulations~\citep{kes_audit:2009}.

%
\acknowledgments
The authors thank the SOHO/MDI consortia for their data. SOHO is a joint project by ESA and NASA. This research was supported by a grant from the ``Ulysses - Ireland-France Exchange Scheme'' operated by the Royal Irish Academy and the Minist\`ere des Affaires Etrang\`eres. PAC is an IRCSET Government of Ireland Scholar.  RTJ is funded by a Marie Curie International European Fellowship under FP6.

\appendix

\bibliographystyle{aa}
\bibliography{wtmm} 

\end{document}